\documentclass[useAMS,usenatbib]{mn2e}
\usepackage[english]{babel}

\usepackage{hyperref}
\usepackage{fancyhdr}
\usepackage{amsfonts}
\usepackage{amsmath}
\usepackage{amssymb}
\usepackage{multicol}
\usepackage{layout}
\usepackage{graphicx}
\usepackage{subfigure}

\usepackage{times}
\usepackage{natbib}

\newif\ifAMStwofonts
\AMStwofontstrue

\newcommand{\aap}{A\&A}
\newcommand{\apj}{ApJ}
\newcommand{\apjl}{ApJL}
\newcommand{\apjs}{ApJS}
\newcommand{\mnras}{MNRAS}
\newcommand{\physrep}{Physics Reports}
\newcommand{\prd}{Phys. Rev. D}

\graphicspath{{./fig/}}

\title[A numerical study of the effects of primordial non-Gaussianities on weak lensing statistics] {A numerical study of the effects of primordial non-Gaussianities on weak lensing statistics}

\author[F. Pace et al.]
       {F. Pace$^{1}$\thanks{E-mail: francesco@ita.uni-heidelberg.de},
         L. Moscardini$^{2,3}$, M. Bartelmann$^{1}$, E. Branchini$^{4}$, K. Dolag$^{5}$, M. Grossi$^{5,6}$, \newauthor
         S. Matarrese$^{7,8}$\\
         $^{1}$ Zentrum f\"ur Astronomie der Universit\"at Heidelberg, Institut f\"ur Theoretische Astrophysik, Albert Ueberle Str. 2, D-69120 Heidelberg, Germany \\
         $^{2}$ Dipartimento di Astronomia, Universit\`a di Bologna, Via Ranzani 1, I-40127 Bologna, Italy \\
         $^{3}$ INFN, Sezione di Bologna, Viale Berti Pichat 6/2, I-40127 Bologna, Italy\\
         $^{4}$ Dipartimento di Fisica, Universit\`a di Roma Tre, Via della Vasca Navale 81, I-00146 Roma, Italy \\
         $^{5}$ Max-Planck-Institut f\"ur Astrophysik, Karl-Schwarzschild Stra\ss e 1, D-85748 Garching, Germany \\
         $^{6}$ Universit\"ats-Sternwarte M\"unchen, Scheinerstrasse 1, D-81679 M\"unchen, Germany\\
         $^{7}$ Dipartimento di Fisica, Universit\`a di Padova, Via Marzolo 8, I-35131 Padova, Italy \\
         $^{8}$ INFN, Sezione di Padova, Via Marzolo 8, I-35131 Padova, Italy \\}

\date{Received \today; accepted ?}

\pagerange{\pageref{firstpage}--\pageref{lastpage}} \pubyear{2010}

\begin{document}
\label{firstpage}
\maketitle

\begin{abstract}
While usually cosmological initial conditions are assumed to be Gaussian, inflationary theories can predict a certain amount of primordial non-Gaussianity which can have an impact on the statistical properties of the lensing observables. In order to evaluate this effect, we build a large set of realistic maps of different lensing quantities starting from light-cones extracted from large dark-matter only N-body simulations with initial conditions corresponding to different levels of primordial local non-Gaussianity strength $f_{\rm NL}$. Considering various statistical quantities (PDF, power spectrum, shear in aperture, skewness and bispectrum) we find that the effect produced by the presence of primordial non-Gaussianity is relatively small, being of the order of few per cent for values of $|f_{\rm NL}|$ compatible with the present CMB constraints and reaching at most 10-15 per cent for the most extreme cases with $|f_{\rm NL}|=1000$. We also discuss the degeneracy of this effect with the uncertainties due to the power spectrum normalization $\sigma_8$ and matter density parameter $\Omega_{\rm m}$, finding that an error in the determination of $\sigma_8$ ($\Omega_{\rm m}$) of about 3 (10) per cent gives differences comparable with non-Gaussian models having $f_{\rm NL}=\pm 1000$. These results suggest that the possible presence of an amount of primordial non-Gaussianity corresponding to $|f_{\rm NL}|=100$ is not hampering a robust determination of the main cosmological parameters in present and future weak lensing surveys, while a positive detection of deviations from the Gaussian hypothesis is possible only breaking the degeneracy with other cosmological parameters and using data from deep surveys covering a large fraction of the sky.
\end{abstract}

\begin{keywords}
cosmology: theory - gravitational lensing: weak - cosmological parameters - large-scale structure of the Universe - Methods: N-body simulations 
\end{keywords}

\section{Introduction}

In recent years, the interest for an accurate measurement of the amount of non-Gaussianity present in the primordial density field has largely increased. The main reason is that this test is now considered not only a general probe of the inflationary paradigm, but also a powerful tool to constrain the plethora of its different variants. Only the most standard slow-rolling models based on a single field produce in fact almost uncorrelated fluctuations, which is the motivation of the common assumption (and large simplification) that their distribution is Gaussian. In general, small deviations from Gaussianity are predicted even for the simplest inflationary models, while non-standard models, like the scenarios based on the curvaton, on the inhomogeneous reheating and on the Dirac-Born-Infeld inflation allow much more significant departures \citep[see][and references therein]{Bartolo2004}.

It has become common to quantify the level of primordial non-Gaussianity adopting the dimensionless non-linearity parameter $f_{\rm NL}$ \citep[see, e.g.,][]{Salopek1990,Gangui1994,Verde2000,Komatsu2001}, that measures the importance of the quadratic term in a sort of Taylor expansion of the gauge-invariant Bardeen potential $\Phi$\footnote{We recall that on scales smaller than the Hubble radius $\Phi$ corresponds to the usual Newtonian peculiar potential (but with changed sign).}:
\begin{equation}
 \Phi=\Phi_{\rm L}+f_{\rm NL} (\Phi^2_{\rm L}-\langle \Phi^2_{\rm L}\rangle)\ ;
\end{equation}
here $\Phi_{\rm L}$ represents a Gaussian random field. Hereafter we will adopt the so-called large-scale structure (LSS) convention, where $\Phi$ is linearly extrapolated to the present epoch\footnote{With the
cosmological parameters adopted in this paper, this corresponds to values for $f_{\rm NL}$ larger by a factor of $\approx 1.3$ with respect to the so-called CMB convention, where $\Phi$ is instead extrapolated at $z=\infty$.}. Moreover we will consider only the so-called local shape for non-Gaussianity, in which the bispectrum signal is larger on squeezed triangle configurations. For more details about other possible shapes we refer to \cite{Bartolo2004,Verde2009,Bartolo2009.1} and references therein.

At present the most stringent constraints on $f_{\rm NL}$ come from the cosmic microwave background (CMB) data. Their discriminating power derives from the fact that its temperature fluctuations trace the
density perturbations before the gravitational non-linearities modify their original distribution. Whatever is the specific test adopted, all CMB analysis consistently allow only very small deviations from
Gaussianity: for instance, analyzing the recent WMAP data, \cite{Komatsu2010} found that $f_{\rm NL}$ varies between -13 and 97, while \cite{Smith2009} found $-5<f_{\rm NL}<104$ (at 95 per cent confidence level); see also \cite{Komatsu2009}.

Alternative and complementary constraints on $f_{\rm NL}$ can in principle be derived analyzing the LSS \citep[for recent reviews, see, e.g.][]{Verde2010,Desjacques2010}. As already evident from the first
generation of non-Gaussian N-body simulations \citep{Messina1990,Moscardini1991,Weinberg1992}, the presence of a positive (negative) skewness in the PDF of the primordial density field tends to favor (disfavor) the formation of cosmic structures, inducing a different timing in the whole process of gravitational instability. However, to be fully exploited, this approach needs to be complemented with reliable methods to disentangle from the primordial signal the non-Gaussian features introduced by the late non-linear evolution and by the possible presence of a non-linear bias factor. While lensing statistics entirely avoid the latter problem, to attack the former one it is necessary to make use of both analytic techniques \citep[like high-order perturbation theory and the Time-Renormalization Group approach; see, e.g.,][]{Taruya2008,Bartolo2009.2} and full N-body simulations \citep[see, e.g.,][]{Kang2007,Grossi2007,Dalal2008,Viel2009,Desjacques2009,Pillepich2010,Grossi2009.1} to properly calibrate the theoretical predictions in the non-linear regime. The large amount of theoretical work recently done in this direction allowed to better understand what is the size of the effects on the abundance of non-linear structures \citep{Matarrese2000,Verde2000,Mathis2004,Kang2007,Grossi2007,Grossi2009.1,Maggiore2009,Roncarelli2010}, on the halo biasing \citep{Dalal2008,McDonald2008,Fedeli2009.1,Carbone2010}, on the galaxy bispectrum \citep{Sefusatti2007.1,Jeong2009,Nishimichi2009.2}, on the mass density distribution \citep{Grossi2008,Lam2009.1,Lam2009.2}, on the topology \citep{Matsubara2003.1,Hikage2008}, on the integrated Sachs-Wolfe effect \citep{Afshordi2008,Carbone2008}, on the Ly-$\alpha$ flux from low-density intergalactic medium \citep{Viel2009}, on the 21 cm fluctuations \citep{Cooray2006,Pillepich2007} and on the reionization process \citep{Crociani2009}. The first attempts of an application to real observational data gave very encouraging results: \cite{Slosar2008}, combining the bias measurements for two samples of luminous red galaxies and quasars, found $f_{\rm NL}=48^{+55}_{-74}$; \cite{Afshordi2008}, studying the integrated Sachs-Wolfe effect (ISW) in the NVSS survey, derived $f_{\rm NL}=354\pm 165$; all error bars are at 2-$\sigma$ level\footnote{We report the values as revised by \cite{Grossi2009.1} to include a correction mimicking the ellipsoidal collapse and converted to the LSS convention.}. Very recently \cite{Xia2010} found $32 < f_{\rm NL}< 152$ from the analysis of the auto-correlation of the brightest NVSS sources on angular scales of several degrees.

In this paper we will focus on estimating the weak lensing signals in scenarios with primordial non-Gaussianity. Being based on the measurement of the shear effect produced by the intervening large-scale structure of the Universe on the images of background galaxies, gravitational lensing is a direct probe of the total matter distribution. For this reason it is considered one of the most powerful tools to constrain the main cosmological parameters and many dedicated projects are in progress or under study. A very exciting perspective is certainly opened by the ESA Cosmic Vision project EUCLID \citep{Laureijs2009}, currently under study: the goal of its wide survey is to obtain the shear measurements for about 40 galaxies per arcmin$^2$ on the entire extragalactic sky with Galactic latitude $b>30$ (approximately 20000 deg$^2$). The possibility of using the weak lensing signals to constrain also the amount of primordial non-Gaussianity has been already explored by different authors. \cite{Amara2004} used a generalized halo model to study the impact on the estimates of the power spectrum normalization $\sigma_8$ of primordial non-Gaussianity, modeled assuming various lognormal distributions for the density field. More recently, \cite{Fedeli2010} computed the power spectrum of the weak cosmic shear for non-Gaussian models with different values of $f_{\rm NL}$. In particular, they improved the halo model including more accurate prescriptions for its ingredients (mass functions, bias and halo profile), calibrated on the last generation of non-Gaussian N-body simulations. The application of this model to a survey having the expected characteristics of the EUCLID project showed the possibility of a significant detection of non-Gaussianity at the level of $|f_{\rm NL}|\approx$ few tens, once the remaining parameters are held fixed.

In this paper we investigate weak lensing statistics in non-Gaussian scenarios using numerical rather than analytical tools. Specifically, we will create weak lensing maps performing ray-tracing simulations through very deep light-cones extracted from high-resolution N-body simulations. The advantage of this approach is twofold. First of all, N-body simulations permit to fully account for the non-linear evolution which is usually modeled less accurately by analytical means. Second, numerical experiments allow us to extract a large set of realistic weak lensing maps that can be used for better evaluating the statistical robustness of the results.

The main goal of our numerical work is to figure out what are the observational evidences of the presence of some level of primordial non-Gaussianity, as quantified by the $f_{\rm NL}$ parameter. In particular we will compute a large set of weak lensing statistics in models with different $f_{\rm NL}$ and we will quantify the deviations from the corresponding results in the Gaussian scenario. This is important not only to address the possibility of a positive detection with future data, but also to establish at which level an amount of primordial non-Gaussianity compatible with the present observational constraints can hamper an accurate measurement of the other cosmological parameters. We must recall that Gaussian initial conditions are virtually always assumed in their practical derivation.

The plan of this paper is as follows. In Section~\ref{sect:lensing} we review the basis of the lensing formalism necessary to the present work. In Section~\ref{sect:simulations} we describe the cosmological
simulations and the numerical procedure to build the lensing maps. In Section~\ref{sect:results} we present our main results about the statistical properties of the different lensing quantities investigated: the probability distribution function, the third-order moment, the power spectrum and the bispectrum. In Section~\ref{sect:params} we compare the effects produced by primordial non-Gaussianity to the uncertainties related to power spectrum normalization $\sigma_8$ and on the matter density parameter $\Omega_{\rm m}$. Finally, in Sect.~\ref{sect:conclusions} we draw our conclusions.

\section{Lensing theory}\label{sect:lensing}

In this section we give a short summary of the aspects of the theory of gravitational lensing that will be used throughout this work. For more detail we refer to the review by \cite{Bartelmann2001}.

We can start by describing how light rays are deflected by the presence of structures in the universe. Since the coherence length of cosmic structures is small compared to the Hubble radius, it is possible to slice the large-scale structures into shells and use the so-called \emph{thin-screen} approximation for them, which allows to consider only the projected mass distribution of the slices. Denoting by $\Sigma(\vec{\theta})$ the projected mass distribution of the lens at the angular position vector $\vec{\theta}$, the convergence can be defined as $\kappa(\vec{\theta})\equiv\Sigma(\vec{\theta})/\Sigma_{\rm
  crit}$, where
\begin{equation}
  \Sigma_{\rm crit}\equiv \frac{c^2}{4\pi G}\frac{D_{ds}}{D_dD_s}
\end{equation}
is the {\em critical surface density}. In the previous equation $D_s$, $D_d$ and $D_{ds}$ represent the angular-diameter distances between the observer and the source, between the observer and the lens, and
between the lens and the source, respectively.

Thanks to the \emph{thin-screen} approximation, the object acting as lens can be completely described using its lensing potential $\Psi$, which is related to the convergence through the two-dimensional Poisson equation
\begin{equation}
  \nabla^2\Psi(\vec{\theta})=2\kappa(\vec{\theta})\;.
\end{equation}
The deflection angle $\hat{\alpha}$ is simply the gradient of the lensing potential, $\hat{\alpha}=\nabla \Psi$.

Up to second order, lensing-induced image distortions are given by
\begin{equation}
  {\theta_i}'\simeq
  A_{ij}{\theta_j}+\frac{1}{2}D_{ijk}\theta_j\theta_k
\end{equation}
\citep[see, e.g.,][]{Goldberg2005,Bacon2006}, where $A_{ij}\equiv \partial_j \theta_i'$ represent the elements of the Jacobian matrix of the lens equation, ${\theta_i}'$ are the unlensed coordinates, and the tensor is defined as $D_{ijk}\equiv \partial_kA_{ij}$. In the previous equations $\partial_i\equiv \partial/\partial\theta_i$. The quantities $A$ and $D$ can be conveniently expressed as a function of the convergence $\kappa$, of the complex shear term $\gamma=\gamma_1+i\gamma_2$, and of their derivatives:
\begin{eqnarray}
  A & = & \left( \begin{array}{cc}
    1-\kappa-\gamma_1 & -\gamma_2         \\
    -\gamma_2         & 1-\kappa+\gamma_1 \\
  \end{array} \right) \nonumber\\
  D_{ij1} & = & \left( \begin{array}{cc}
    -2\gamma_{1,1}-\gamma_{2,2} & -\gamma_{2,1} \\
    -\gamma_{2,1}               & -\gamma_{2,2} \\
  \end{array} \right) \\
  D_{ij2} & = & \left( \begin{array}{cc}
    -\gamma_{2,1} & -\gamma_{2,2}              \\
    -\gamma_{2,2} & 2\gamma_{1,2}-\gamma_{2,1} \\
  \end{array} \right) \;;\nonumber
\end{eqnarray}
in the previous equations the comma indicates the derivative. The shear derivatives can be combined to construct two new quantities, $F$ and $G$, called {\em first} and {\em second flexion} respectively, defined as
\begin{eqnarray}
  F & \equiv & F_1+iF_2=(\gamma_{1,1}+\gamma_{2,2})+i(\gamma_{2,1}-\gamma_{1,2}) \\
  G & \equiv & G_1+iG_2=(\gamma_{1,1}-\gamma_{2,2})+i(\gamma_{2,1}+\gamma_{1,2}).
\end{eqnarray}

The previous formalism can be easily generalized to the case in which a continuous distribution of matter is considered. The volume between the observer and the sources can be divided in a sequence of sub-volumes having a size along the line-of-sight sufficiently small compared to the distances between the observer and the sub-volumes, and between those and the sources. The matter of each sub-volume can be projected onto a plane and then we are allowed to use again the thin-screen approximation described above. The final quantities, estimated on the source plane, will be the weighted sum of the relevant quantities, where the weight is given by a suitable ratio of the involved distances. In multiple lens-plane theory, rotation of light bundles can occur in addition to shear and convergence, but numerical simulations have shown that these are negligibly small \citep[]{Jain2000}. Then, all the lens properties are contained in the lensing potential. Knowing it on all lens planes allows us to write recursion relations representing the deflection angle, the shear, the effective convergence and the two flexions on the source plane. These relations will be given and discussed in more detail in Section~\ref{sect:raytracing}.

\section{The lensing simulations}\label{sect:simulations}
\subsection{The cosmological simulations}

To study the effect of non-Gaussianity on the weak lensing statistics, we use the outputs of a set of N-body cosmological simulations following the evolution of dark matter particles only. These simulations have been already used by \cite{Grossi2007}, \cite{Hikage2008}, \cite{Grossi2008} and \cite{Roncarelli2010}. Here we will summarize the information relevant for this paper, referring the interested readers to the original papers for further details.

The simulations were carried out using the publicly available code {\small GADGET-2} \citep{Springel2005} assuming a $\Lambda$CDM model with total matter density $\Omega_{\rm m}=0.3$, baryon density
$\Omega_{\rm b}=0.04$ and cosmological constant density $\Omega_{\Lambda}=0.7$; the Hubble parameter is set to $h=0.7$, while the power spectrum is normalized to $\sigma_8=0.9$. The simulated box has a comoving length of 500 Mpc/h and contains $800^3$ particles with a mass resolution of $2\times 10^{10}~M_{\odot}/h$. The comoving softening scale is $\epsilon_{\rm Pl}=12.5~\rm{kpc/h}$. The different outputs are equispaced in comoving space by 250 Mpc/h. The set is composed by seven cosmological simulations considering different values of the non-Gaussianity parameter $f_{\rm NL}$: $\pm 1000$, $\pm 500$, $\pm 100$, plus the standard Gaussian case $f_{\rm NL}=0$. Notice that in this work we prefer to make use of this set of N-body simulations in spite of the more recent one presented in \cite{Grossi2009.1}. The reason for this choice is twofold. First, the mass resolution is better and this allows us to have more robust results on small scales, in particular for shear and flexions; second, the assumed values for the parameter $f_{\rm NL}$ span a larger range, covering also cases where the weak lensing signal produced by primordial non-Gaussianity is larger.

\subsection{Building the mock light-cones}

To construct the mock light-cones used to perform the ray-tracing simulations, we follow the same procedure described in \cite{Pace2007.1}, to which we refer for more details. Since the different outputs of an N-body simulation represent the redshift evolution of the same initial matter distribution, we need to apply a specific procedure to avoid the introduction of biases related to the fact that the same structures appear approximately at the same positions in different outputs. For this reason we randomly shift and rotate the particle positions exploiting the periodicity of the simulation boxes in the plane perpendicular to the line-of-sight. As said before, given our choice for the output redshifts, they overlap for $50$ per cent of their comoving side-length, therefore we can just consider particles in the lower half of the rotated and translated boxes.

Once selected the particles to be used, we project them along the line-of-sight on a regular two-dimensional grid and compute the projected mass density field using the triangular-shaped-cloud (TSC) mass assignment \citep{Hockney1988}. Finally via fast Fourier transform (FFT) techniques it is possible to recover the lensing potential associated with the considered matter distribution. More details on the numerical procedure will be given in the next section.

Our mock light-cones extend along the line-of-sight up to redshift $z_{\rm s}=1$, using $N_{\rm out}=9$ outputs. As all the distances are in comoving units, the opening angle of the ray-tracing simulation can be calculated using the comoving distance of the last plane in the stack, resulting in $\theta=13.49$ degrees. Using a grid of $2048^2$ points, the corresponding angular resolution of the produced maps is 23.7 arcsec.

\subsection{Ray-tracing simulations}
\label{sect:raytracing}
Ray-tracing simulations consist in tracing back a bundle of light rays through the matter distribution of the light-cone, from the observer to the sources, which we place at redshift $z_{\rm s}=1$.

Projected mass maps are converted to projected density-contrast maps $\delta^{{\rm proj,i}}_{\rm lm}$ by
\begin{equation}
  \delta^{{\rm proj,i}}_{\rm lm}=\frac{M^{\rm i}_{\rm lm}}{A_{\rm i}\bar\rho}-L_{\rm i} \;,
\end{equation}
where $A_i$ is the the area of the grid cell on the $i$-th plane, $M^{\rm i}_{\rm lm}$ is the mass projected on the grid cell with indices $(l,m)$ belonging to the $i$-th plane and $L_{\rm i}=250$ Mpc/h is the depth of the $i$-th subvolume used to build mock cone. The density contrast is deliberately defined such as to have the unit of a length \citep[see][]{Hamana2001}. In the previous equation $\bar\rho$ represents the average comoving density of the Universe.

The lensing potential on each plane $\Psi_i$ is related to the projected density contrast through the two-dimensional Poisson equation, namely:
\begin{equation}\label{eq:poisson}
  \nabla^2\Psi_i(\vec{\theta})=3\Omega_{\rm m}\left(\frac{H_0}{c}\right)^2
  \delta^{{\rm proj,i}}(\vec{\theta})\;.
\end{equation}
Exploiting the periodic boundary conditions of the projected maps, Eq.~(\ref{eq:poisson}) can be solved adopting the FFT techniques. Having the potential on each plane, the lensing quantities can be derived adopting standard finite-difference schemes.

In order to perform the ray-tracing simulations, we need to apply the multiple-plane theory to compute the total effect taking into account the contributions from each single lensing plane. A light-ray is
deflected on each plane by the amount $\vec\alpha_i(\vec\theta_i)$, thus the total deflection is given by the sum of all contributions. In particular, if the light-cone is sampled into $N_{\rm out}$ lens planes and the sources are located on the $N_{\rm out}+1$ plane, the relation giving the deflection angle on the $i$-th plane of a ray with image position $\vec{\theta}_1$ reads (in comoving units)
\begin{equation}\label{eq:defangle}
  \vec{\theta}_i = \vec{\theta}_1-\sum_{k=1}^{i-1}\,\frac{f_K(w_i-w_k)}
      {f_K(w_i)a_k}\nabla_{\vec{x}}\Psi_k(\vec{x})\;;
\end{equation}
here $\Psi_k(\vec{x})$ is the unscaled lensing potential, i.e. the Newtonian potential projected along the line-of-sight, $f_K$ is a function depending on the cosmology, $w$ is the comoving distance and $a$ represents the scale factor of the lens plane. Note that in general the light-rays will intercept the lens planes at arbitrary points, while the potential is defined only at the grid points. Thus, it is necessary to use a bi-linear interpolation to compute the lensing quantities.

Differentiating Eq.~(\ref{eq:defangle}) with respect to $\vec{\theta}_1$ and defining $A_i\equiv \partial\vec{\theta}_i/\partial\vec{\theta}_1$ and $U_k$ the matrix containing the second derivatives of the lensing potential, one obtains
\begin{equation}\label{eq:jac}
  A_i=I-\sum_{k=1}^{i-1}\frac{f_K(w_k)f_K(w_i-w_k)}{f_K(w_i)a_k}U_kA_k\;.
\end{equation}
In the previous equation, $I$ represents the identity matrix. On the source plane, the Jacobian matrix $A_{N_{\rm out}+1}$ is given by
\begin{equation}\label{eq:Jacobian}
  A_{N_{\rm out}+1}=\left( \begin{array}{cc}
    1-\kappa-\gamma_1 & -\gamma_2+\omega \\
    -\gamma_2-\omega  & 1-\kappa+\gamma_1
    \end{array}\right)\;,
\end{equation}
where $\kappa$ is now the effective convergence and $\gamma=\gamma_1+i\gamma_2$ is the effective shear. The term $\omega$, called rotation, represents the asymmetry introduced by multiple lenses.

Differentiating Eq.~(\ref{eq:jac}) with respect to $\vec{\theta}_i$, a recursive relation for the two flexions can be obtained:
\begin{equation}\label{eq:hessian}
  D_i^{1,2}=-\sum_{k=1}^{i-1}\frac{f_K(w_k)f_K(w_i-w_k)}{f_K(w_i)a_k}
  [f_k{w_k}G_U^{1,2}+U_kD_k^{1,2}]\;,
\end{equation}
where $G_U=\nabla_{\vec{x}}U$ is a tensor containing the third derivatives of the lensing potential.

On the source plane, the tensor $D$ reads as
\begin{eqnarray}
\label{eq:Dmatrix}
  D_{N+1}^1 & = & \left( \begin{array}{cc}
    -2\gamma_{1,1}-\gamma_{2,2} & -\gamma_{2,1}+\omega_1 \\
    -\gamma_{2,1}-\omega_1               & -\gamma_{2,2} \\
  \end{array} \right)\\
  D_{N+1}^2 & = & \left( \begin{array}{cc}
    -\gamma_{2,1} & -\gamma_{2,2}+\omega_2              \\
    -\gamma_{2,2}-\omega_2 & 2\gamma_{1,2}-\gamma_{2,1} \\
  \end{array} \right) \nonumber\;,
\end{eqnarray}
where, as for the Jacobian matrix, $\omega_1$ and $\omega_2$ represent the asymmetric terms. In Eq.~(\ref{eq:Dmatrix}) we do not take into account the additional terms, called twist and turn, introduced by
\cite{Bacon2009}, but several tests performed at different resolutions assure us that their inclusion would not significantly affect our results.

\section{Results}\label{sect:results}
In this section we present the results of our analysis of the statistical properties of several lensing quantities extracted from the ray-tracing simulations described above. In particular we discuss the probability distribution function (PDF) in Sect.~\ref{sect:PDF}, the third-order moment (skewness) in Sect.~\ref{sect:moments}, the power spectrum in Sect.~\ref{sect:PS}, the shear in aperture in Sect.~\ref{sect:aperture} and finally the bispectrum in Sect.~\ref{sect:biSpec}.

\subsection{Probability distribution functions}
\label{sect:PDF}

\begin{figure*}
  \begin{center}
    \includegraphics[angle=-90,width=0.33\hsize]{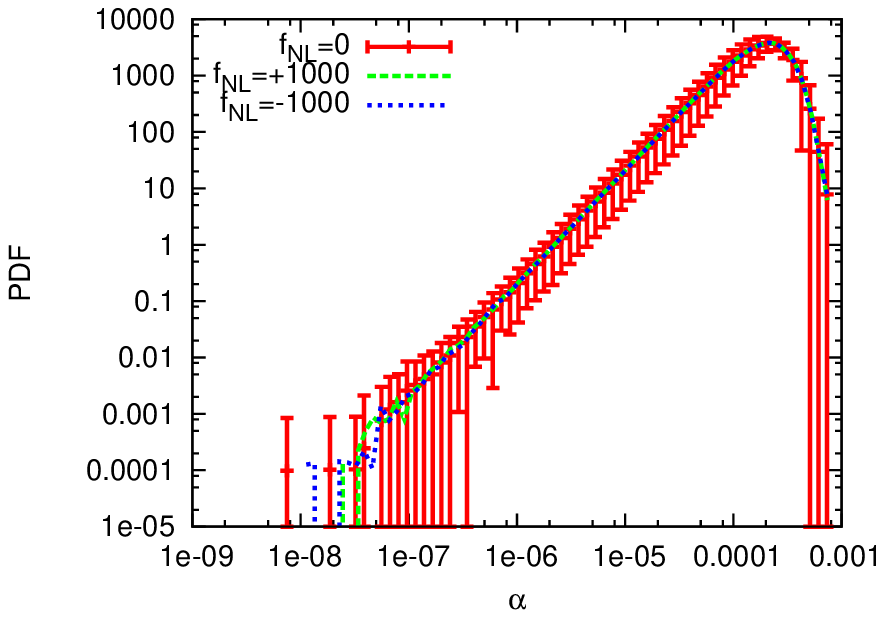}
    \includegraphics[angle=-90,width=0.33\hsize]{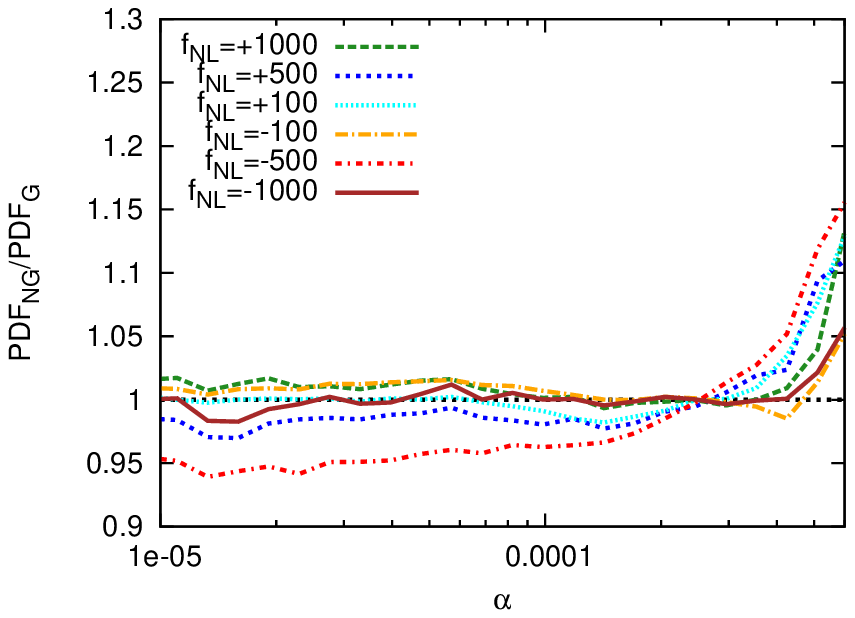}\\
    \includegraphics[angle=-90,width=0.33\hsize]{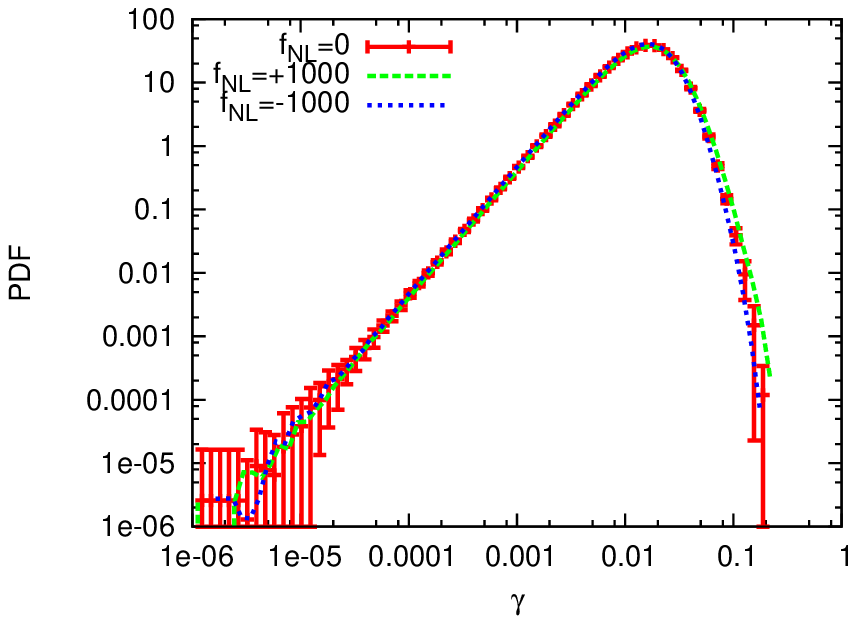}
    \includegraphics[angle=-90,width=0.33\hsize]{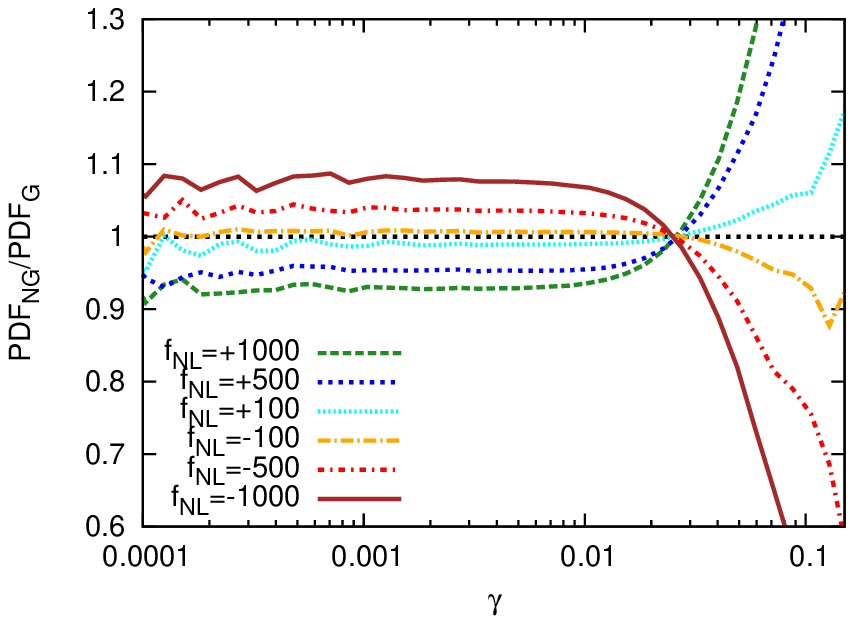}\\
    \includegraphics[angle=-90,width=0.33\hsize]{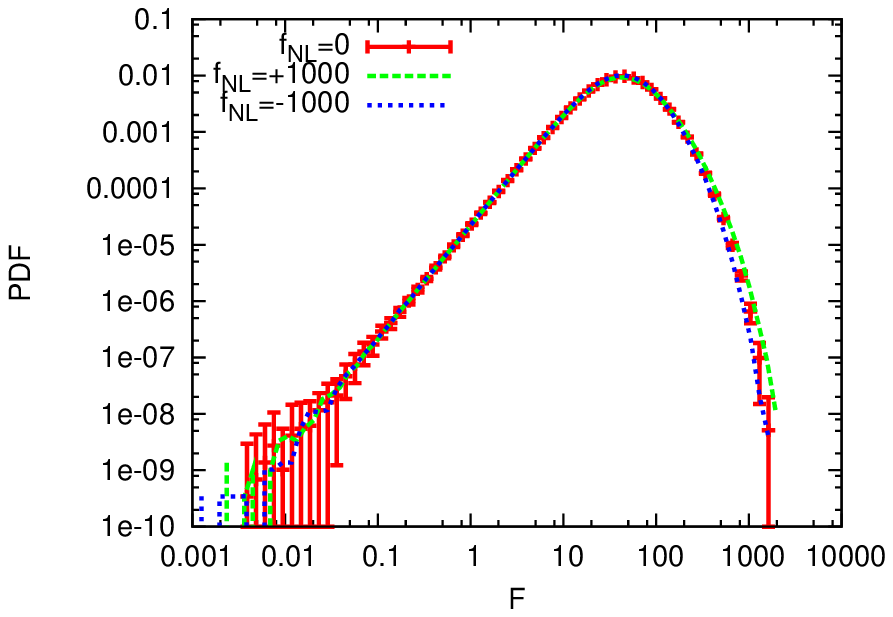}
    \includegraphics[angle=-90,width=0.33\hsize]{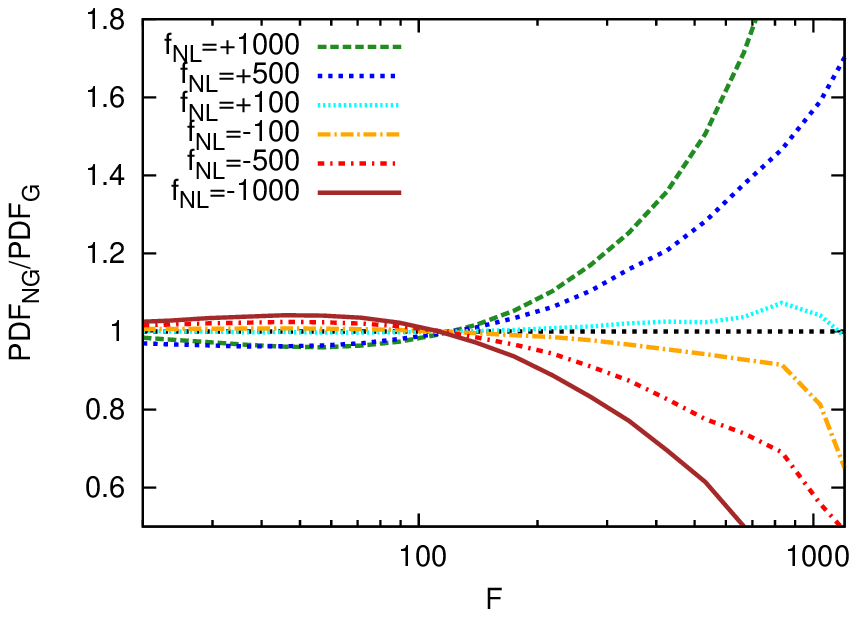}\\
    \includegraphics[angle=-90,width=0.33\hsize]{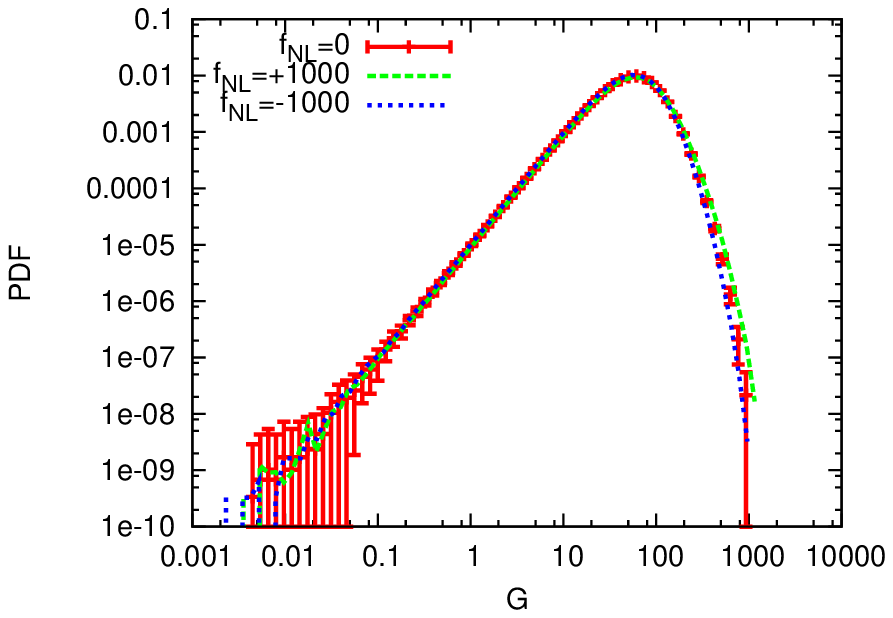}
    \includegraphics[angle=-90,width=0.33\hsize]{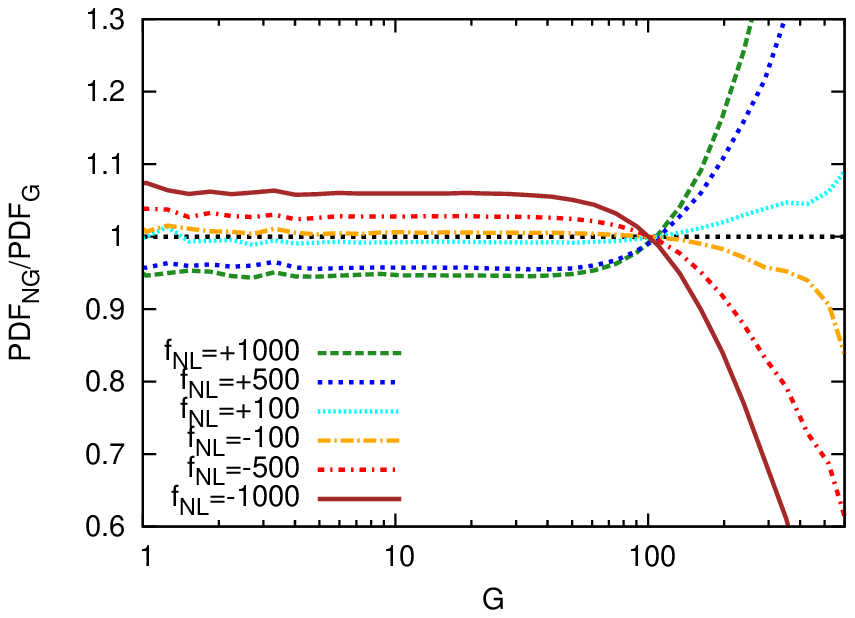}\\
    \includegraphics[angle=-90,width=0.33\hsize]{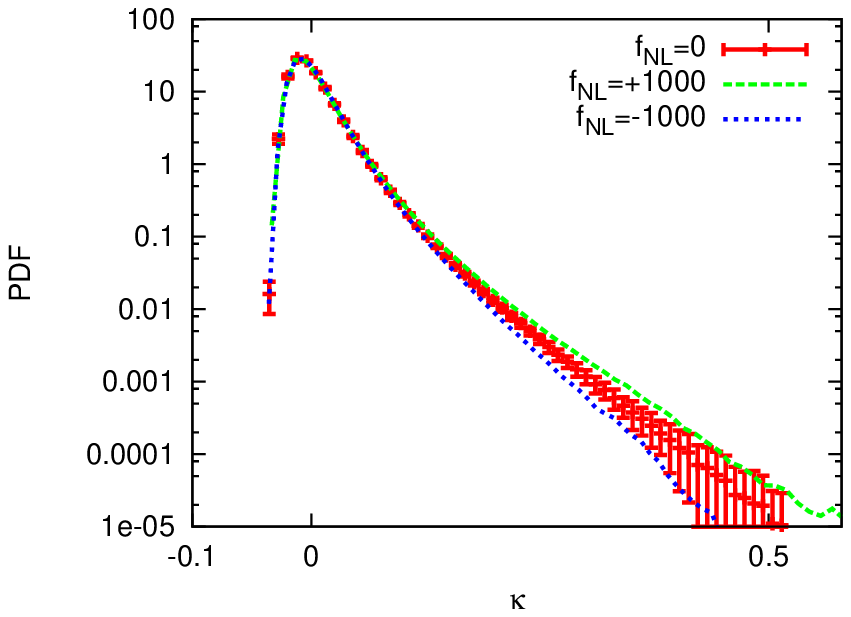}
    \includegraphics[angle=-90,width=0.33\hsize]{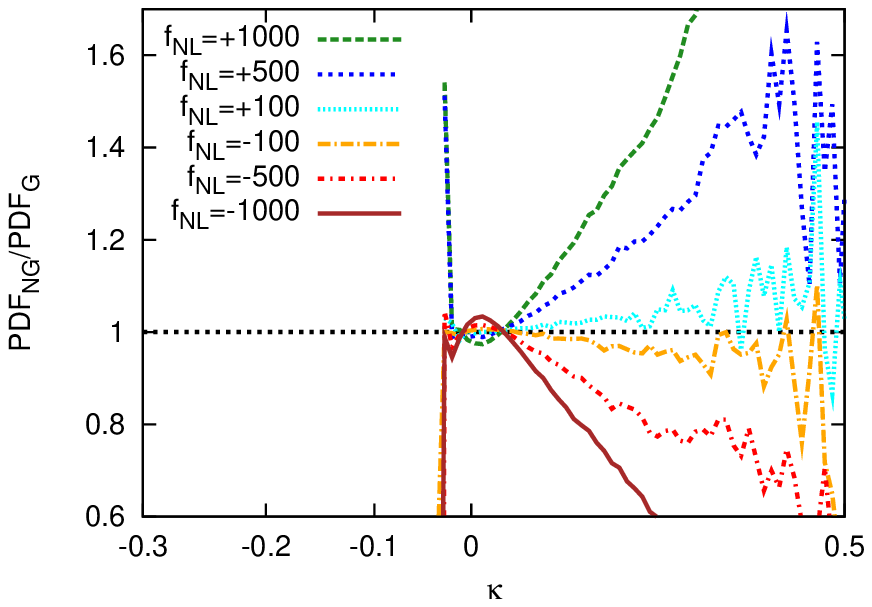}
  \end{center}
  \caption{Left panels: the differential PDF for different lensing quantities: the (modulus of the) deflection angle $\alpha$ (first row); the (modulus of the) shear $\gamma$ (second row); the (moduli of the) two flexions $F$ and $G$ (third and fourth rows, respectively); the effective convergence $\kappa$ (last row). Points and error bars represent the average and r.m.s. of a set of 60 different realizations for the Gaussian model, while blue and red lines present the average values of the simulation set corresponding to the two most extreme non-Gaussian models, $f_{\rm NL}=-1000$ and $f_{\rm NL}=+1000$ respectively. Right panels: the ratio between non-Gaussian and Gaussian differential PDFs for the same quantities shown in the left panels. Different color lines refer to models with different levels of primordial non-Gaussianity: $f_{\rm NL}=+1000$ (green), $f_{\rm NL}=+500$ (blue), $f_{\rm NL}=+100$ (cyan), $f_{\rm NL}=-100$ (orange), $f_{\rm NL}=-500$ (red) and $f_{\rm NL}=-1000$ (brown).}
  \label{fig:PDF}
\end{figure*}

We start considering the PDFs for several important lensing quantities. All our results are grouped in Fig.~\ref{fig:PDF}. The panels on the left present the actual PDFs of a given quantity, as obtained averaging over 60 different light-cone realizations. Given the smallness of the differences, we prefer to show only the results for the Gaussian model $f_{\rm NL}=0$ (red points and lines) and for the most extreme non-Gaussian models: $f_{\rm NL}=-1000$ (blue lines) and $f_{\rm NL}=+1000$ (green lines): models with smaller non-Gaussianity are intermediate between the displayed lines. For reference, we also show (for the Gaussian model only) the error bars, referring to the r.m.s. of the corresponding sample of realizations. Finally, the panels on the right present for all models (represented by different colour lines, as described in the label) the ratio between the non-Gaussian and Gaussian results.

The top row of Fig.~\ref{fig:PDF} refers to the modulus of the deflection angle $\hat \alpha$. The presence of primordial non-Gaussianity affects the PDFs by at most few percent in the most extreme cases with $f_{\rm NL}=\pm 1000$, while the models with primordial non-Gaussianity more consistent with current constraints ($f_{\rm NL}=\pm 100$) are almost indistinguishable from the Gaussian case. We also notice that the differences produced by non-Gaussianity are more evident in the high-value tails. These large deflections reflect lensing events associated with rare, large structures. In this sense these results confirm those on the halo abundance: unlikely events (large haloes) are useful probes of primordial non-Gaussianity. However, we have to note that in the high-value tails the statistics can be quite poor and noisy: a good modelization of the strong non-linear effects acting on the same scales is necessary to allow a positive detection.

In the panels in the second row of Fig.~\ref{fig:PDF} we show the analogous plots for the (modulus of the) shear $\gamma$. Since a negative value of the $f_{\rm NL}$ parameter favors lower values of $\gamma$, we find that the ratio between non-Gaussian and Gaussian models is in this case larger than unity; the opposite trend holds for models with positive primordial non-Gaussianity. As already pointed out for the deflection angle, the effect of a mild non-Gaussianity on the shear distribution is tiny, below 1 per cent, which is comparable to the size of the error bars (shown only in the left panels) but can grow up to 10 per cent for the very extreme tail. Only models with very high values of $f_{\rm NL}$ display a deviation with respect to the Gaussian case that can be as large as the error bars obtained by averaging over the simulated maps.

The PDFs for the first and second flexion are shown in the panels in the third and forth rows of Fig.~\ref{fig:PDF}, respectively. The trend is very similar to the case of the shear: compared to the Gaussian model, models with negative $f_{\rm NL}$ have a higher probability of an excess at low values; the opposite trend holds in the high-value tail. The differences between Gaussian and non-Gaussian models are slightly more evident for the second flexion $G$.

Finally the panels in the last row of Fig.~\ref{fig:PDF} refer to the effective convergence, for which we find a slightly different situation. In this case, in the maps there are also negative values, corresponding to underdense regions where the non-Gaussian PDFs show large deviations from the Gaussian results: for negative $\kappa$ we find less than 1 per cent for $f_{\rm NL}=\pm 100$, $\approx 1$ per cent for $f_{\rm NL}=\pm 500$, and approximately 3 per cent for $f_{\rm NL}=\pm 1000$. Note that this result agrees with the analysis made by \cite{Grossi2008} on the three-dimensional density field and its probability distribution: underdense regions tend to maintain the imprint of primordial non-Gaussianity, suggesting that statistics based on voids can be a powerful tool to estimate $f_{\rm NL}$ \citep[see also][]{Kamionkowski2009}. The plot shows that also the overdense regions keep the important imprints of the primordial non-Gaussianity: for $\kappa=0.3$ we find 8-10 per cent for $f_{\rm NL}=\pm 100$, $\approx 40$ per cent for $f_{\rm NL}=\pm 500$, and more than 80 per cent for $f_{\rm NL}=\pm 1000$. However, in this regime, the statistics is quite poor and the ratio between non-Gaussian and Gaussian PDFs becomes very noisy. As final comment on PDFs (and on their moments, see the following sections), it is important to stress that the lensing quantities here discussed are observables that directly measure the dark matter distribution. This is not true for the density PDF derived from galaxy surveys, where the complex effect of bias must be corrected for.

\subsection{Third-order moment}
\label{sect:moments}

A possible alternative way to detect the signatures produced by the presence of some primordial non-Gaussianity is to look at the high-order moments (skewness, kurtosis, etc.) of the distribution of the various lensing quantities. The power of this approach was already evident in the investigation of the corresponding quantities related to the density field \citep[see, e.g.,][]{Coles1993,Grossi2008,Lam2009.1}. However, the measurement of these statistics is often affected by large error bars, that increase with the order and hamper their reliable application to the real data. For this reason, in this section we will focus only on the skewness $\mu_3$ of the two-dimensional convergence field $\kappa_{i,j}$, defined as
\begin{equation}
  \mu_3 = \frac{\mu_2^{-3/2}}{N^2} \sum_{i,j}(\kappa_{i,j}-\bar{\kappa})^3\ ,
\end{equation}
where $\bar{\kappa}$ and $\mu_2 \equiv \sum_{i,j}(\kappa_{i,j}-\bar{\kappa})^2/{N^2}$ represent the mean and the variance, computed on all $N^2$ pixels of the maps. To derive these quantities, we first convolve the convergence map with a Gaussian filter of angular radius $\theta$, then we subtract the mean and finally we normalize it appropriately.

\begin{figure}
  \begin{center}
   \includegraphics[angle=-90,width=0.8\hsize]{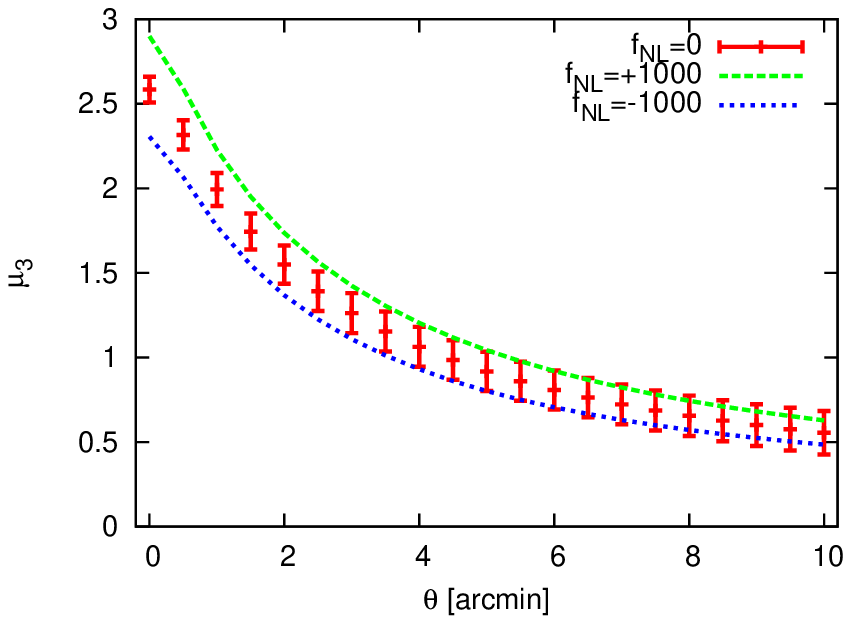}\hfill
   \includegraphics[angle=-90,width=0.8\hsize]{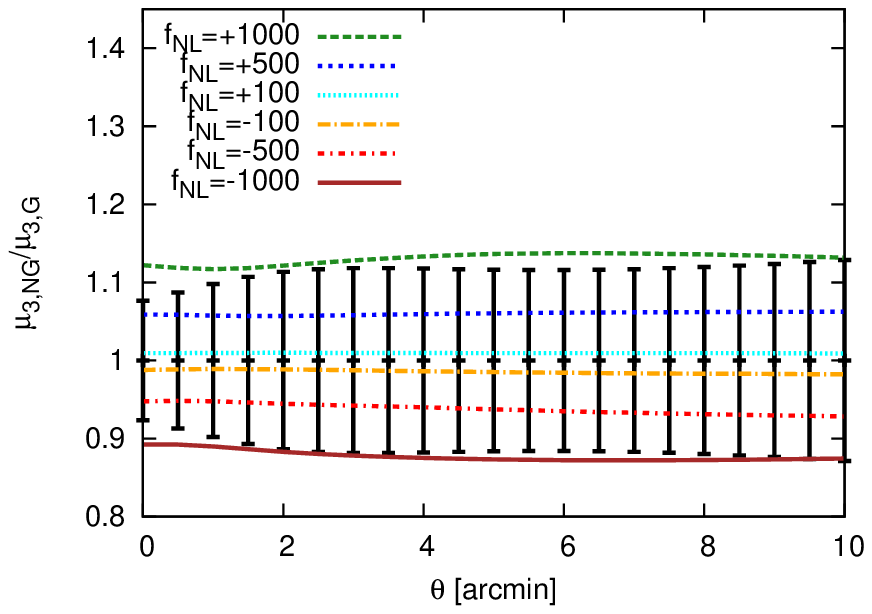}
  \end{center}
  \caption{Upper panel: the skewness $\mu_3$ of the effective convergence for the Gaussian model as a function of the angular scale $\theta$. The data and error bars represent the mean and the r.m.s. computed on a set of 60 different light-cone realizations, while blue and red lines present the average values of the simulation set corresponding to the two most extreme non-Gaussian models, $f_{\rm NL}=-1000$ and $f_{\rm NL}=+1000$ respectively. Lower panel: the ratio between the skewness results for non-Gaussian and Gaussian models. Different color lines refer to different values of $f_{\rm NL}$, as indicated in the labels.}
  \label{fig:Mkappa}
\end{figure}

In the upper panel of Fig.~\ref{fig:Mkappa} we present the results for the skewness of the effective convergence, as extracted from the Gaussian simulation. For reference we also show the results for the two most extreme non-Gaussian models. As expected, $\mu_3$ (which is computed as average over 60 different realizations) is a decreasing function of the filtering radius $\theta$: increasing the smoothing reduces the non-Gaussian features introduced by the non-linear evolution. Furthermore the error bars, representing the r.m.s. over the set of different light-cone realizations, are slightly increasing with $\theta$ due to increasing shot noise, since by increasing the smoothing radius we are averaging over a smaller number of circles. This behavior is analogous to what found by \cite{Jain2000} in the determination of the skewness of the effective convergence.

More interesting is the bottom panel of the same figure, where we show the ratio between the third-order moments computed in non-Gaussian and Gaussian simulations: we find a variation of the order of 12 per cent for $f_{\rm NL}=\pm 1000$, $\approx 6$ per cent for $f_{\rm NL}=\pm 500$ and only $\approx 0.8$ per cent for $f_{\rm NL}=\pm 100$. It is worth noticing that this effect is almost constant on scales up to 10 arcmin. Comparing these results with the size of the error bars, we can conclude that with relatively small maps, like those analyzed in this paper, we can use the skewness statistics to detect only strong primordial non-Gaussianities ($f_{\rm NL}=\pm 500, \pm 1000$). As shown by the results of the analysis made by \cite{Fedeli2010}, only with data covering very large area it would be possible to disentangle the effects of non-Gaussian models with $f_{\rm NL}=\pm 100$. Finally we notice that a similar analysis performed on the shear field, which is directly related to the convergence one, would provide comparable results.

\subsection{Power Spectrum}
\label{sect:PS}

An important theoretical quantity, directly related to the matter power spectrum, is the effective convergence power spectrum $P_{\kappa}(\ell)$, which is defined as the squared modulus of the Fourier transform of $\kappa$, averaged on the modes having a given multipole $\ell$. Starting from $P_{\kappa}(\ell)$, it is possible to derive analytic expressions for the power spectra of the other lensing quantities, such as those for the shear ($P_{\gamma}$) and the two flexions ($P_{F}$ and $P_{G}$), namely:
\begin{eqnarray}
  \label{eq:lps2}
  P_{\gamma}(\ell) & = & P_{\kappa}(\ell) \\
  \label{eq:lps3}
  P_{F}(\ell) & = &  P_{G}(\ell) = \ell^2P_{\kappa}(\ell) \ .
\end{eqnarray}

\begin{figure}
  \begin{center}
    \includegraphics[angle=-90,width=0.45\hsize]{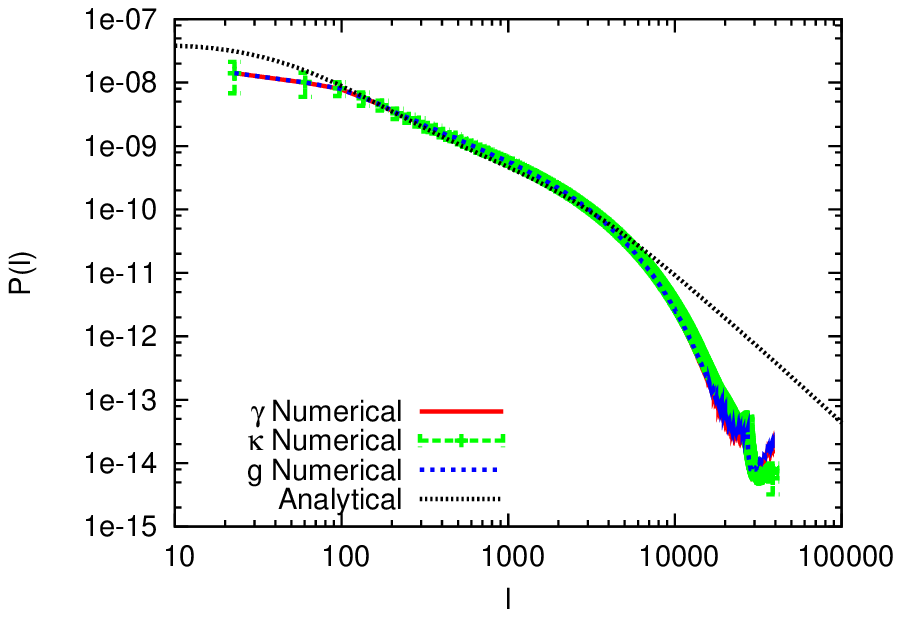}
    \includegraphics[angle=-90,width=0.45\hsize]{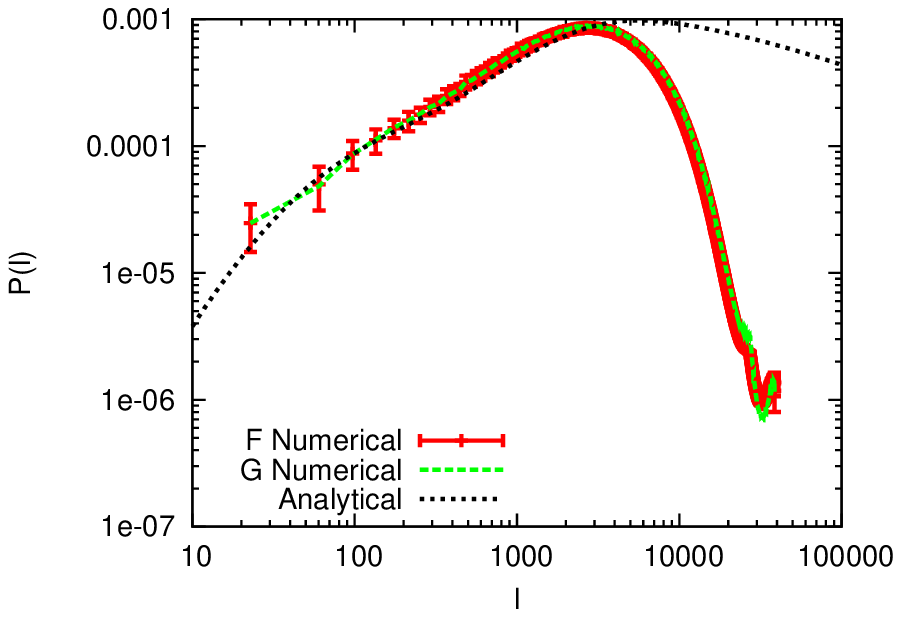}\\
    \includegraphics[angle=-90,width=0.45\hsize]{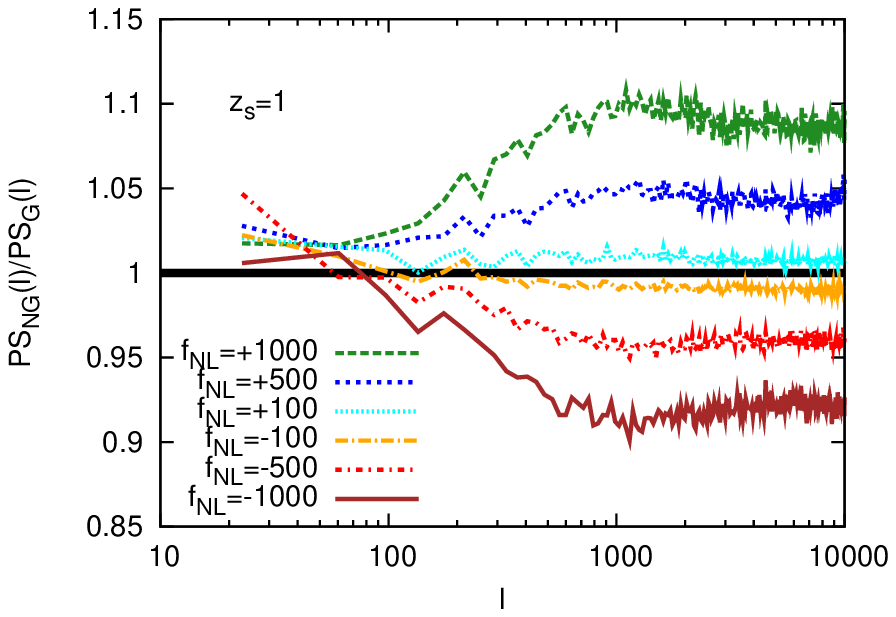}
    \includegraphics[angle=-90,width=0.45\hsize]{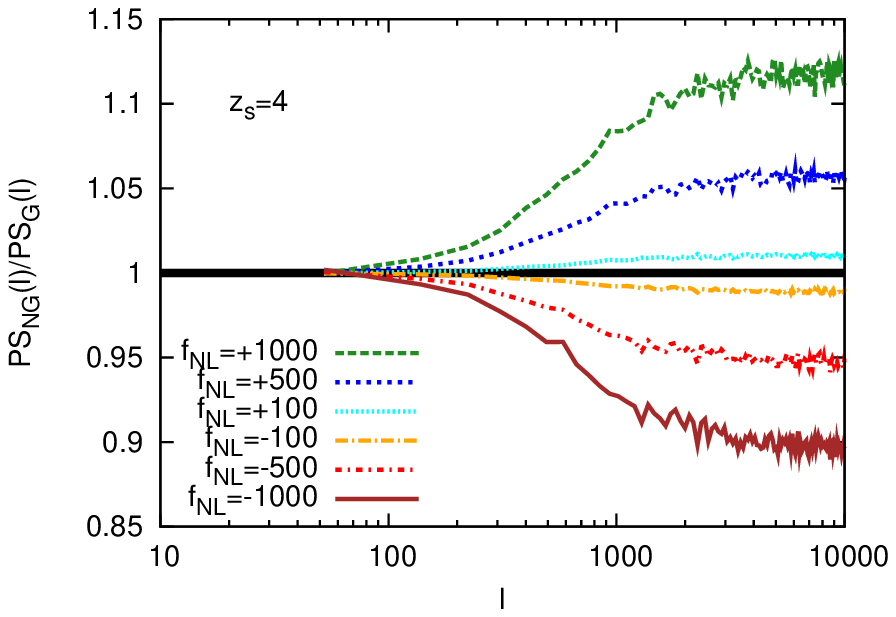}
  \end{center}
  \caption{Upper panels: comparison between theoretical predictions and our numerical estimates, derived as average of 60 different realizations of the power spectra. Error bars correspond to the r.m.s. over the different realizations. Various lensing quantities are shown: effective convergence, shear and reduced shear (upper left panel); first and second flexions (upper right panel). The results refer to the Gaussian case. Bottom panels: ratio between non-Gaussian and Gaussian estimates for the convergence power spectrum (averaged over 60 different realizations) as a function of the multipole $l$. Different color lines refer to various values of $f_{\rm NL}$ as indicated in the labels. The left (right) panel refers to sources at $z_{\rm s}=1$ ($z_{\rm s}=4$).}
  \label{fig:PS}
\end{figure}

In the top panels of Fig.~\ref{fig:PS} we show, for the Gaussian model only, the comparison between the power spectra extracted from our simulations and the corresponding theoretical predictions including the non-linear effect. We consider different lensing quantities: effective convergence, shear and reduced shear $g\equiv\gamma/(1-\kappa)$ (upper left panel) and two components of the flexion (top right panel). We notice that the agreement between the numerical results and the theoretical expectations holds, as expected, only in a given range of wavenumbers. On large scales (small $\ell$), the departure from theory is due to the low number of modes we can use to average the numerical power spectra. On small scales (large $\ell$), the disagreement is produced by the mass resolution (particularly relevant for the flexion power spectrum) and by numerical artifacts produced by the weighting function appearing in the definition of a given power spectrum. A better mass resolution will make the lack of power at high frequencies less severe, but it will not solve it entirely because of the limit of resolution due to the Nyquist frequency intrinsic to every discrete Fourier transformation. Moreover, when numerically performing the ray-tracing simulations, low-redshift planes are poorly sampled since a few low-redshift pixels contribute to the lensing maps. This poor sampling is amplified by the weighting function appearing in Eqs.~\ref{eq:defangle},~\ref{eq:jac} and ~\ref{eq:hessian}. We can use Fig.~\ref{fig:PS} as a reference to define the range of $\ell$ on which the numerical results can be trusted: typically $ 10^2 \la \ell\la 10^4$.

In the bottom panels of Fig.~\ref{fig:PS} we show the ratio between the power spectra derived from the non-Gaussian and the Gaussian simulations. We present here only the results for the effective convergence since the other power spectra can be obtained by introducing the suitable dependence on $\ell$, and then the corresponding ratios are identical. In this case, we decided to study the effect of primordial non-Gaussianity at two different source redshifts, namely $z_{\rm s}=1$ and $z_{\rm s}=4$. This is done to verify, with the help of numerical simulations, recent analytical predictions made by \cite{Fedeli2010} using an improved version of the halo model. As expected, the effects of the primordial non-Gaussianity become more evident when the absolute value of the parameter $f_{\rm NL}$ increases. At very large scales, where the evolution is still in the linear regime, the Gaussian model and all the non-Gaussian ones here considered display approximately the same power spectrum, and the ratio is in practice unity or slightly larger due to the scatter of the power spectrum at low $\ell$. At smaller scales ($\ell>100-200$) the behaviors start to be different: a positive (negative) $f_{\rm NL}$ parameter implies a larger (smaller) power spectrum with respect to the Gaussian case. It is also worth noticing that at very small scales ($\ell>4000-5000$) the ratio tends to an asymptotic value: this is due to the fact that in this range the power spectrum can be very well approximated by a power-law relation and all models turn out to have approximately the same slope. We have to recall, however, that some numerical artifacts can affect our results on these small scales. Looking at the plots, we see that, as expected, the differences between Gaussian and non-Gaussian models are tiny, for the power spectrum too and increase progressively with the source redshift: for the most extreme models ($f_{\rm NL}=\pm 1000$), the deviations are of the order of 8 per cent at $z_{\rm s}=1$ (10-12 per cent at $z_{\rm s}=4$), for the models with $f_{\rm NL}=\pm 500$ they are approximately 4-5 per cent, while for the models with $f_{\rm NL}=\pm 100$ the deviations are smaller than 1 per cent. From the bottom panels of Fig.~\ref{fig:PS} we also see that the peak of the ratio between the non-Gaussian and the Gaussian models shifts at higher frequencies when the source redshift increases. This is in agreement with what was found recently by \cite{Fedeli2010}.

\subsection{Shear in aperture}
\label{sect:aperture}

A quantity which is directly related to the shear power spectrum but which can be more easily measured in real data is the so-called {\it shear in aperture} $\gamma_{\rm av}$. It represents the variance of the shear field within a circular aperture of a given radius $\theta$:

\begin{equation}
  |\gamma_{\rm av}(\theta)|^2\equiv 2\pi\int_0^{\infty}~dl~lP_{\gamma}(l)
  \left[\frac{J_1(l\theta)}{\pi l\theta}\right]^2 \ ,
\end{equation}
where $J_1$ is the first-order Bessel function of the first kind.

\begin{figure}
 \begin{center}
  \includegraphics[angle=-90,width=0.8\hsize]{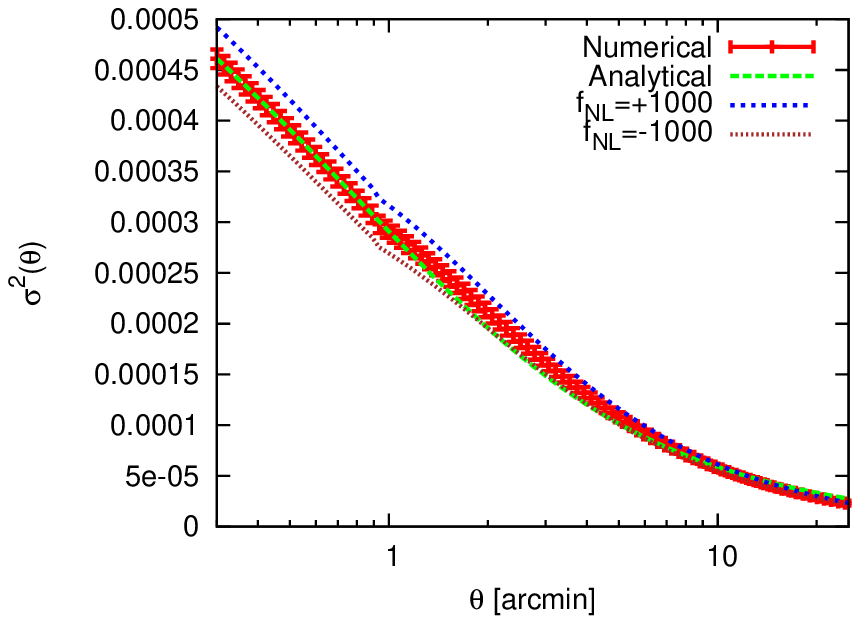}
  \includegraphics[angle=-90,width=0.8\hsize]{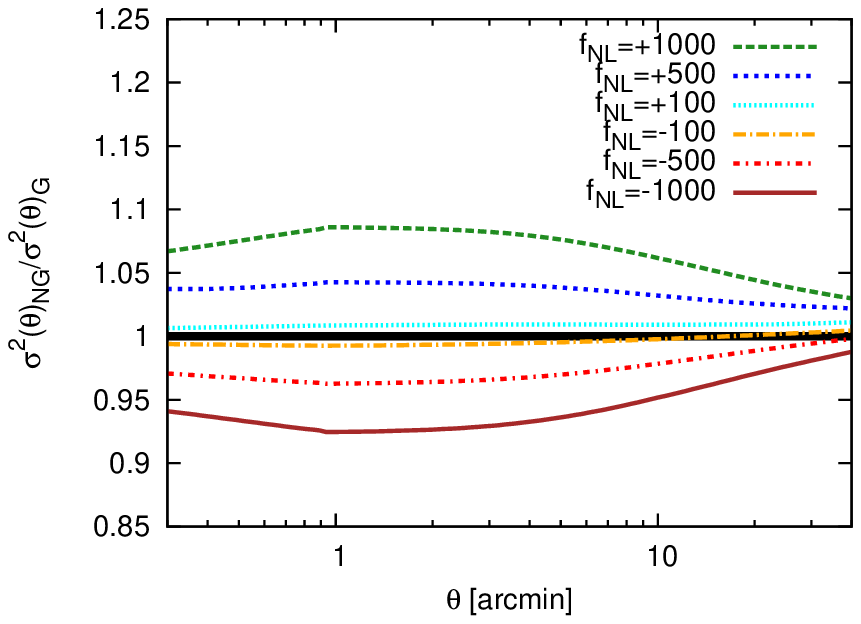}
 \end{center}
 \caption{The shear in aperture. Top panel: comparison between the theoretical analytical relation and our estimate, derived as average of 60 numerical realizations; error bars represent the r.m.s. The results refer to the Gaussian model. Blue and brown lines present the corresponding results for the two most extreme non-Gaussian models, $f_{\rm NL}=+1000$ and $f_{\rm NL}=-1000$ respectively. Bottom panel: the ratio between the results for non-Gaussian and Gaussian models: different color lines refer to various values of $f_{\rm NL}$ as indicated in the labels.}
 \label{fig:SVgamma}
\end{figure}

The top panel of Fig.~\ref{fig:SVgamma} shows the comparison between the theoretical expectation and the results derived from our Gaussian simulations (averaged over 60 different light-cone realizations): the
good agreement is evident up to scales of $\theta \approx 20$ arcmin. For reference we also plot the results for $f_{\rm NL}=\pm 1000$. The bottom panel of the same figure presents the ratio between the results for non-Gaussian and Gaussian models as a function of the aperture angle. Being a weighted integral of the shear power spectrum, $|\gamma_{\rm av}(\theta)|^2$ has a trend which is quite similar to the one displayed by $P_\gamma$: the larger the absolute value of the $f_{\rm NL}$ parameter is, the larger are the deviations from the Gaussian expectations. An increment (decrement) in the signal of the order of 7-8 per cent is expected for the model with $f_{\rm NL}=+1000$ ($f_{\rm NL}=-1000$), while the differences reduce to less than 1 per cent for the models with $f_{\rm NL}=\pm 100$. As expected, we find that the deviations from the Gaussian expectations are larger for small apertures, but rapidly wash out for angles larger than few arcmin, since the average of the shear involves more and more large scale structures. We also notice that the ratio at large smoothing scales does not reach unity since structures are already evolved enough to show high intrinsic non-Gaussianity, therefore higher smoothing scales are necessary to compensate the effect.

\subsection{Bispectrum}
\label{sect:biSpec}
In several works \citep[see e.g.][]{Verde1998,Verde2000.1,Takada2004,Sefusatti2006,Sefusatti2010} it has been shown that a large amount of cosmological information can be inferred from the study of high-order spectra (the so-called poly-spectra). Here we just limit our investigation to the simplest high-order spectrum, the bispectrum. We recall that for a Gaussian random field all available information is contained in the power spectrum, while all higher-order spectra vanish or are combinations of the power spectrum. As a consequence the measurement of the bispectrum is quite sensitive to the non-Gaussian features of the quantity under investigation.

In general, the bispectrum depends on a triangular configuration and its values are very sensitive to the particular configuration adopted. Since the complete determination of the bispectrum is extremely expensive from a computational point of view, we will restrict our analysis to equilateral triangles only. Notice that, given our assumption of local shape for the primordial non-Gaussianity, the non-Gaussian signal on squeezed configurations is expected to be larger. In a future work we will use the convergence maps presented in this work to study the full bispectrum configuration and how the signal expected from primordial non-Gaussianity is sensitive to the triangular configuration.

In general the bispectrum of a given quantity $f$ is defined as:
\begin{equation}
  \langle\hat{f}(\vec{\ell}_1)\hat{f}(\vec{\ell}_2)\hat{f}(\vec{\ell}_3)\rangle=(2\pi)^2\delta_D(\vec{\ell}_{123})B(\vec{\ell}_1,\vec{\ell}_2,\theta_{12})\;.
\end{equation}
In the previous equation, $\hat{f}$ represents the Fourier transform of the field $f$. Since the wavenumbers at which the bispectrum is computed ($\vec{\ell}_1$, $\vec{\ell}_2$ and $\vec{\ell}_3$) must form a closed triangle in Fourier space, they are related by the constraint $\vec{\ell}_{123}=\vec{\ell}_1+\vec{\ell}_2+\vec{\ell}_3=0$. Finally $\theta_{12}$ is the angle between $\vec{\ell}_1$ and $\vec{\ell}_2$
and fixes, together with the triangle condition, $\vec{\ell}_3$.

\begin{figure}
 \begin{center}
   \includegraphics[angle=-90,width=0.8\hsize]{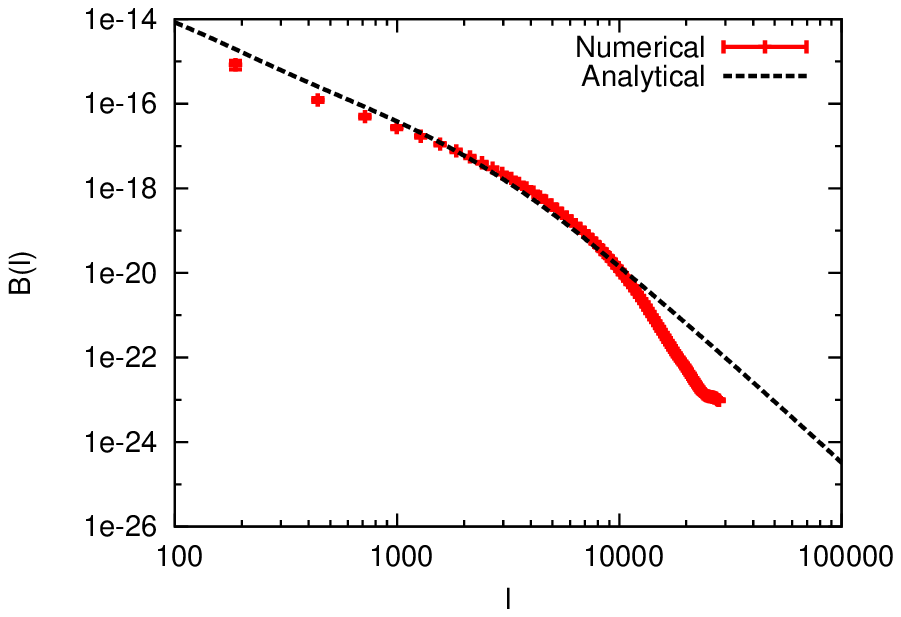}
   \includegraphics[angle=-90,width=0.8\hsize]{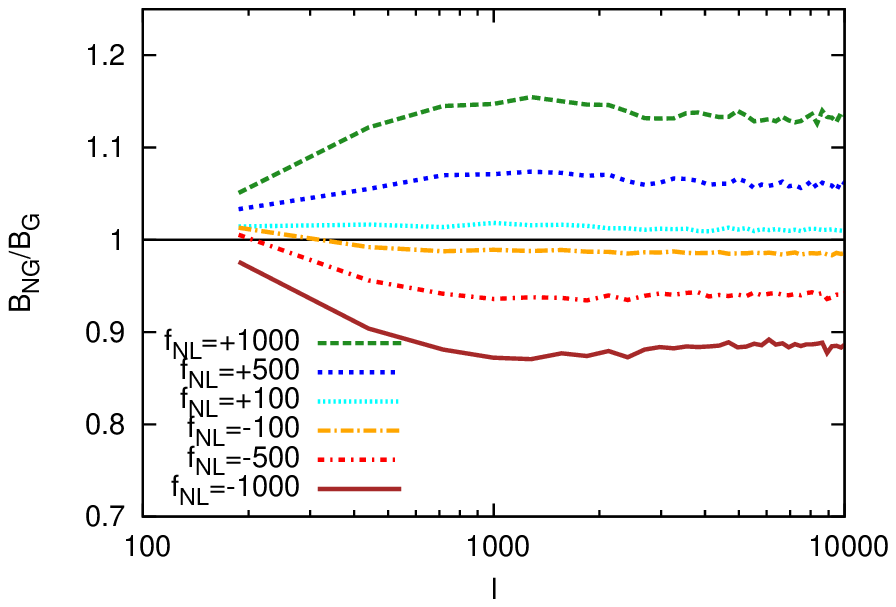}
 \end{center}
 \caption{Top panel: comparison between theoretical predictions and our numerical estimates, derived as average of 60 different realizations of the bispectrum for the effective convergence. Error bars represent the r.m.s. of the 60 different realizations. The results refer to the Gaussian model only. Bottom panel: the ratio between non-Gaussian and Gaussian estimates for the bispectrum as a function of the multipole $\ell$. Different color lines refer to various values of $f_{\rm NL}$ as indicated in the labels.}
 \label{fig:BS}
\end{figure}

Similar to Fig.~\ref{fig:PS}, the top panel of Fig.~\ref{fig:BS} shows, for the Gaussian model only, the comparison between our numerical estimates and the theoretical bispectrum computed for the effective convergence. We show a bispectrum computed for an equilateral configuration, i.e., in the Fourier space, $|\vec{\ell}_1|=|\vec{\ell}_2|=|\vec{\ell}_3|$. The results from the simulations represent the average over 60 different light-cone realizations with the corresponding r.m.s. plotted as error bars. The theoretical quantity are derived from the matter density bispectrum $B_{\rm{PT}}(\vec{k}_1,\vec{k}_2,\theta_{12})$, that can be computed according to the perturbation theory. For instance, in the case of the effective convergence $\kappa$, the bispectrum can be obtained by using the Limber's approximation in an analogous way as for the power spectrum \citep[see e.g.][]{Cooray2001}, namely:
\begin{eqnarray}
  B_{\kappa}(\vec{\ell}_1,\vec{\ell}_2,\theta_{12})&=&\int_0^{w_{\rm lim}}\frac{{\rm d}w}
{f_K(w)}
  G^3(w)\nonumber\\
  &\times& B_{\rm{PT}}\left(\frac{\vec{\ell}_1}{f_K(w)},\frac{\vec{\ell}_2}
{f_K(w)}\right)\ , 
\end{eqnarray}
where $f_K(w)$ is a function depending on the cosmological parameters and $G(w)$ contains the distance weight \citep{Bartelmann2001}. For the other lensing quantities, similar relations hold. In order to quantify the effects of primordial non-Gaussianity, the panel on the bottom shows the bispectrum computed for the different non-Gaussian models, normalized to the corresponding result for the Gaussian one.

As for the case of the power spectrum, we see that the agreement between the theoretical expectations and the numerical predictions holds only for a limited range of angular scales, but in this case the range for which the estimate of bispectrum is robust is much more limited. Focusing on scales where the numerical results can be considered reliable, we find that the contribution of non-Gaussianity is not very relevant, reaching up to 12 per cent in the very non-linear regime, but only for the most extreme non-Gaussian models ($f_{\rm NL}=\pm 1000$). For models with $f_{\rm NL}$ in the range allowed by the most recent CMB constraints, we do not expect a difference with respect to the Gaussian case larger than few per cent.

\section{The degeneracy between primordial non-Gaussianity and the power spectrum normalization and matter density parameter}\label{sect:params}
As shown in the previous sections, the effects due to primordial non-Gaussianity on the various weak lensing quantities are small, in particular when we consider models with values of $f_{\rm NL}$ in the range allowed by present data. To better investigate the possibility of their future detection, here we will discuss their degeneracy with the effects produced by the variation of other relevant cosmological parameters. In particular we will focus on the degeneracy between the amount of primordial non-Gaussianity and the power spectrum normalization $\sigma_8$ and matter density parameter $\Omega_{\rm m}$, which are certainly the most important source of uncertainty.

Since the different timing in the structure formation induced by $f_{\rm NL}$ modifies the power spectrum and its amplitude, non-Gaussianity can be confused with a Gaussian model with a different $\sigma_8$ or $\Omega_{\rm m}$ parameter. In particular a positive (negative) $f_{\rm NL}$ has effects similar to a higher (lower) power spectrum normalization or matter density parameter. In this section we show how much the uncertainty on the value of the $\sigma_8$ normalization affects the possibility of detecting non-Gaussianity, keeping all the other parameters fixed. We also show how important is the error in the determination of the matter density parameter. To do so, we vary at the same time both $\Omega_{\rm m}$ and $\Omega_{\Lambda}$ to have a flat geometry. For the sake of simplicity, we restrict our analysis to the convergence power spectrum, the shear in aperture and the effective convergence bispectrum.

\begin{figure*}
 \begin{center}
   \includegraphics[angle=-90,width=0.33\hsize]{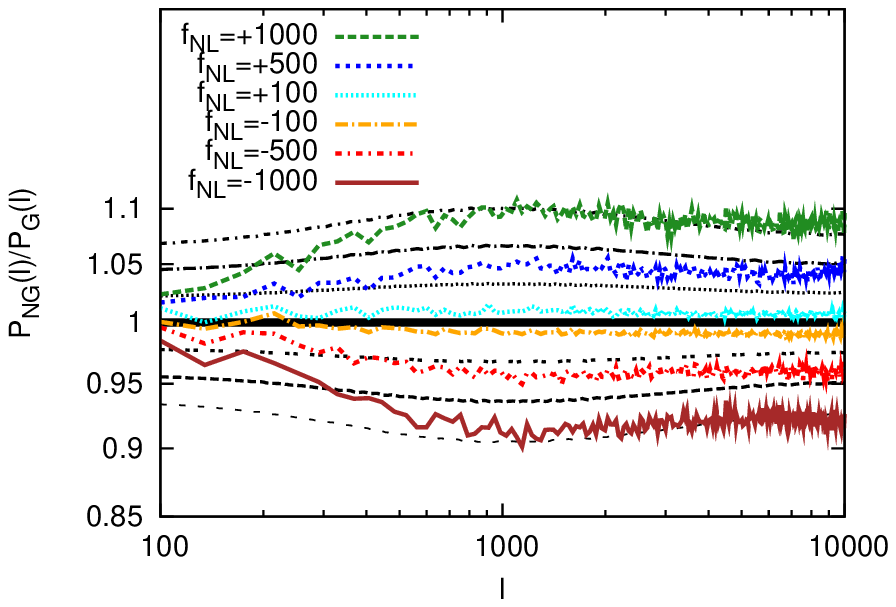}
   \includegraphics[angle=-90,width=0.33\hsize]{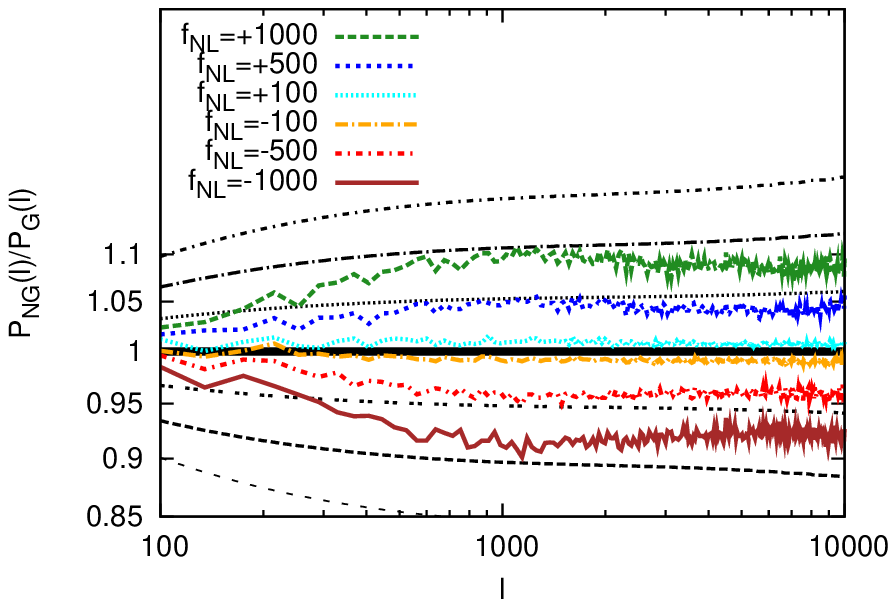}\\
   \includegraphics[angle=-90,width=0.33\hsize]{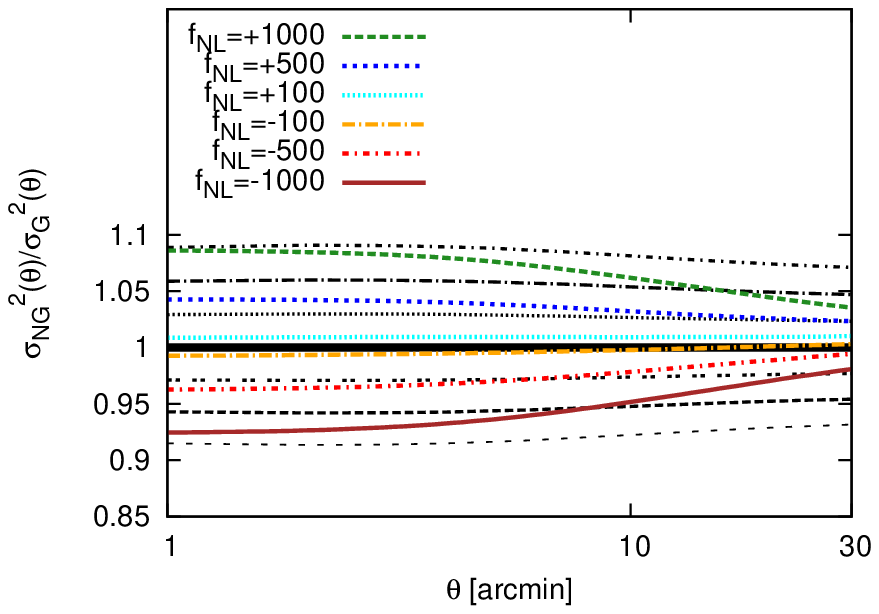}
   \includegraphics[angle=-90,width=0.33\hsize]{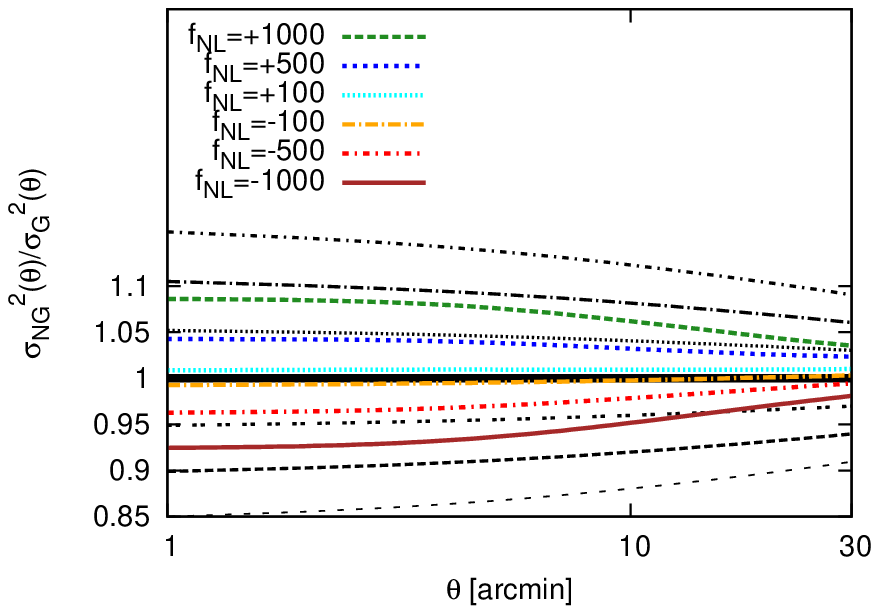}\\
   \includegraphics[angle=-90,width=0.33\hsize]{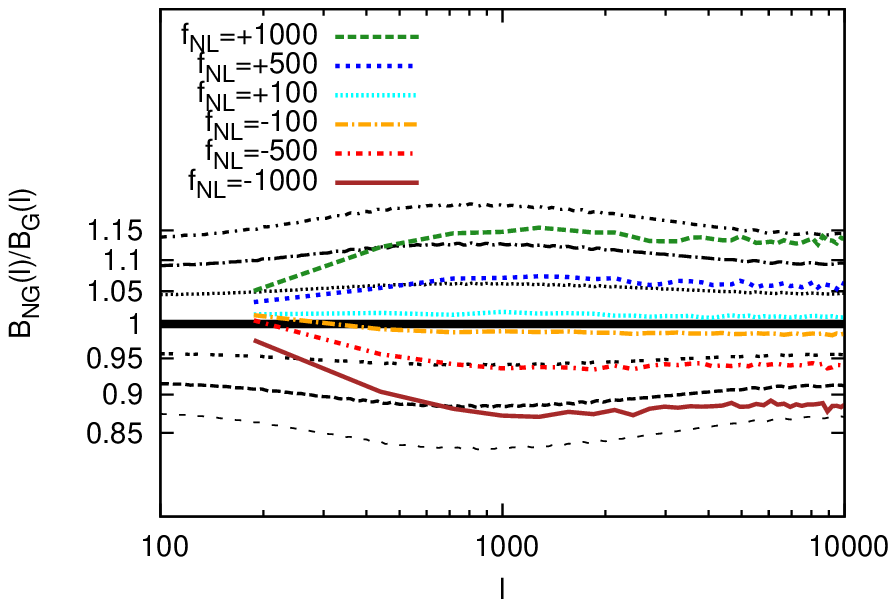}
   \includegraphics[angle=-90,width=0.33\hsize]{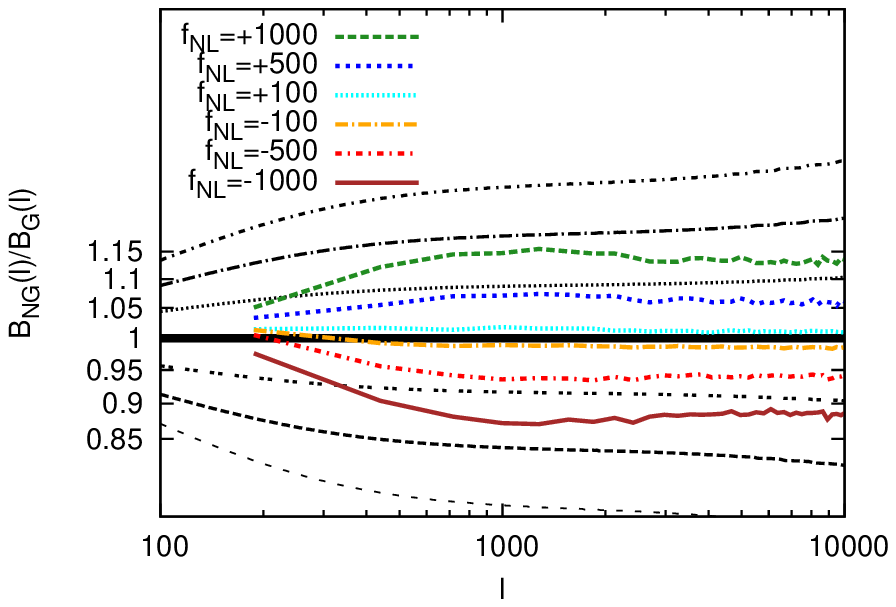}
 \end{center}
 \caption{Comparing the effects of different levels of primordial non-Gaussianity to the ones produced by different power spectrum normalizations (left panels) and matter density parameter (right panels). The panels show various lensing quantities: effective convergence power spectrum (upper panel); shear variance (central panel); effective convergence bispectrum (lower panel). The results are presented as ratio with respect to the same quantity computed for the Gaussian model with $\sigma_8=0.9$ ($\Omega_{\rm m}$=0.3) (solid horizontal line) and refer to Gaussian models with different $\sigma_8$ normalizations  ($\Omega_{\rm m}$ parameter): 0.93 (0.33) (dot-dashed line), 0.92 (0.32) (dot with long dash line), 0.91 (0.31) (dotted line), 0.89 (0.29) (double dots line), 0.88 (0.28) (dashed line with three dots), 0.87  (0.27) (short dashed line). For reference we also show the effects produced by different levels of primordial non-Gaussianity: $f_{\rm NL}=+1000$ (cyan line); $f_{\rm NL}=+500$ (blue line); $f_{\rm NL}=+100$ (red line); $f_{\rm NL}=-100$ (magenta line); $f_{\rm NL}=-500$ (orange line); $f_{\rm NL}=-1000$ (brown line).}
 \label{fig:params}
\end{figure*}

In the left column of Fig.~\ref{fig:params} we show a comparison between the contribution of non-Gaussianity and the changes produced by a variation of the spectrum normalization while on the right panels we show the effect of varying the matter content, keeping $\Omega_{\rm m}+\Omega_{\Lambda}=1$. In particular we consider $\sigma_8$ ($\Omega_{\rm m}$) in the range between 0.87 and 0.93 (0.27 and 0.33): the size of this variation approximately corresponds to the typical 2$\sigma$ error derived from the most recent CMB analysis. It is clear from the figure that the error on the determination of the parameters has effects more important than the introduction of a weak primordial non-Gaussianity. The two most extreme models produce deviations from the Gaussian predictions that are of the same level of uncertainty $\Delta \sigma_8=0.03$ and $\Delta\Omega_{\rm m}=0.02$. On the contrary, the models with $f_{\rm NL}$ in the range allowed by current constraints have effects well within the range allowed by the present uncertainties, since their effect is smaller than a variation of $\sigma_8 (\Omega_{\rm m}) \approx0.01$. This implies that significant constraints on the amount of primordial non-Gaussianity can be obtained only reducing the uncertainties on the power spectrum normalization and on the matter density parameter.

\section{Conclusions}\label{sect:conclusions}

In this work, we used the outputs of N-body simulations to create a large set of realistic mock maps for several lensing quantities (deflection angle, effective convergence, shear and the two components of the flexion) in the framework of cosmological models with different amount of primordial non-Gaussianity, quantified using the dimensionless parameter $f_{\rm NL}$. In particular we considered several statistical properties (PDF, power spectrum, shear in aperture, skewness and bispectrum) and compared the results with the corresponding ones obtained in the Gaussian case. Our main results can be summarized as follows.

\begin{itemize}
\item For all quantities here considered the effect produced by the presence of primordial non-Gaussianity is relatively small, amounting to differences of 1 per cent for $|f_{\rm NL}|=100$, 5 per cent for $|f_{\rm NL}|=500$, and 8-15 per cent for $|f_{\rm NL}|=1000$. These results are in good agreement with the analytic predictions presented in \cite{Fedeli2010}.
\item The largest effects are visible on small scales (i.e. for large multipoles $l>1000$), where, however, also non-linearity can produce strong effects which have to be accurately modeled.
\item The most promising statistical tests to search for imprints of primordial non-Gaussianity are the (convergence and shear) power spectra and the (convergence) bispectrum, thanks to the smaller size of their error bars at the relevant scales.
\item The differences of the various PDFs in both rare-event tails are also important, but their discriminating power is reduced by the poor statistics and by the high-level of noise.
\item We compared the effects produced by the primordial non-Gaussianity with the uncertainties due to the power spectrum normalization and matter density parameter: an error in the determination of $\sigma_8$ of about 3 per cent or of $\Omega_{\rm m}$ of about 10 per cent gives differences comparable with the non-Gaussian models with $f_{\rm NL}=\pm 1000$, while for more realistic non-Gaussian models with $f_{\rm NL}=\pm 100$ the effects is smaller than the one induced by $\Delta\sigma_8 (\Delta\Omega_{\rm m})=0.01$.
\end{itemize}

As said, a significant covariance exists between primordial non-Gaussianity and fundamental cosmological parameters, especially with $\sigma_8$ and $\Omega_{\rm m}$. The smallness of non-Gaussian effects found in our analysis means that one can obtain a precise estimate of these parameters despite their covariance with $f_{\rm NL}$. As an example, ignoring the possible presence of a primordial non-Gaussianity with $f_{\rm NL}=100$, consistent with current observational constraints, would induce a mere 0.2 per cent uncertainty in the estimate of the cosmological parameters, i.e. well below the current 1-$\sigma$ error. The same argument, in reverse, is telling us that the search for non-Gaussian features in the weak-lensing statistics is a very challenging task. According to our results, it can be successfully completed only if errorbars can be significantly reduced. For the purpose of discriminating among competing models the reduction needs not to be dramatic. Indeed, as pointed out by \cite{Fedeli2010}, the fact that deviations from Gaussianity are small but systematics allows one to estimate $f_{\rm NL}$ by adding the statistical information from different angular bins, this is also shown in our analysis of the power spectrum for sources at different redshifts.

A more significant reduction in the errorbars is required to break the degeneracy between non-Gaussianity with cosmological parameters, say $\sigma_8$ or $\Omega_{\rm m}$. Indeed, as shown in Fig.~\ref{fig:params}, removing such degeneracy requires to compare information from a limited number of bins at very different scales. As shown by \cite{Fedeli2010}, both tasks could be achieved with next-generation, all-sky surveys. A good example is certainly represented by the ESA mission EUCLID \citep{Laureijs2009}. According to the analysis made by \cite{Fedeli2010} and based on analytic predictions for the power spectrum quantitatively confirmed in this numerical work, we expect that the quality and the quantity of the EUCLID data will allow to constrain $f_{\rm NL}$ at the level of few tens, opening the possibility of discriminating between the various inflationary models.

\section*{Acknowledgments}
Computations have been performed on the IBM-SP5 at CINECA (Consorzio Interuniversitario del Nord-Est per il Calcolo Automatico), Bologna, with CPU time assigned under an INAF-CINECA grant, and on the IBM-SP4
machine at the ``Rechenzentrum der Max-Planck-Gesellschaft'' at the Max-Planck Institut fuer Plasmaphysik with CPU time assigned to the ``Max-Planck-Institut f\"ur Astrophysik'' and at the ``Leibniz-Rechenzentrum'' with CPU time assigned to the Project ``h0073''. This work was supported by the Deutsche Forschungsgemeinschaft (DFG) under the grants BA 1369/5-1 and 1369/5-2 and through the Transregio Sonderforschungsbereich TR 33, as well as by the DAAD and CRUI through their Vigoni program. We also acknowledge partial financial contributions from the ASI contracts I/016/07/0 COFIS, Euclid-Dune I/064/08/0, ASI-INAF I/023/05/0, ASI-INAF I/088/06/0, ASI/INAF Agreement I/072/09/0 for the Planck LFI Activity of Phase E2 and from the grant ANR-06-JCJC-0141. K.~D. acknowledges the support by the DFG Priority Programme 1177 and additional support by the DFG Cluster of Excellence "Origin and Structure of the Universe". We warmly thank Massimo Meneghetti, Cosimo Fedeli and Melita Carbone for useful discussions and comments. We also thank the anonymous referee whose comments helped us to improve the presentation of our results.

\bibliographystyle{mn2e}

\label{lastpage}

\end{document}